# ASSESSING THE STOCHASTIC PROPERTIES OF MODERN PSEUDO-RANDOM GENERATORS FOR PARALLEL COMPUTING

Théau Wartel, David R.C. Hill
Université Clermont Auvergne, Clermont Auvergne INP, ENSM St Etienne,
ISIMA/LIMOS UMR CNRS 6158
F-63000 Clermont–Ferrand, FRANCE
E-mail: David.Hill@uca.fr

## Abstract

Pseudo-random number generators (PRNGs) are widely used in modern computing and are expected to exhibit excellent statistical performance and repeatability. This study evaluates and compares modern PRNGs used in high performance computing and artificial intelligence. Our selections comes from different families, including Xoshiro, Philox, PCG, and MRG32k3a. We systematically assess the quality of these generators; instead of testing a single stream for each generator, we test more than $10^3$ streams with the BigCrush battery form the TestU01 library. The results, involving more than 4.5 years of cumulative computing time, are analyzed against the claims made by the generators' creators. The highest success rate is 72%, and all tests have been failed by almost every generator, the failed tests are documented. To ensure fairness, all tests are conducted under consistent conditions and are designed to closely simulate real-world usage. The results of each test are available, usable and reproducible with a git repository.

## Introduction

Pseudorandom number generators (PRNGs) are widely used in various fields of modern computer science: numerical simulation of physical models, Monte Carlo simulations, artificial intelligence (AI) and machine learning. To obtain high-quality results, the generators on which our results are based must be of excellent quality. The quality of a generator depends on various points: the numbers are generated uniformly, the generator is portable on different platforms, the sequence can be changed (with a different initialization state) and is repeatable (Antunes and Hill, 2024), the period is as large as possible, the generator uses little memory and numbers are obtained quickly, the generator produce uncorrelated sequences which pass common statistical tests. The ability to be easily parallelized is also important for intensive computing (Hill et al. 2013).

In this paper, we focus on the statistical quality of the generators, i.e. their ability to generate a sequence of numbers with the lowest possible correlation between numbers. This has to be achieved in a perfectly deterministic way to ensure debugging as well as the repeatability of numerical experiences. The fact that we only evaluate the statistical quality of PRNGs means that our results do not reflect their overall performance. A generator that may seem inadequate according to this criterion could nonetheless be relevant for other usage. In our study, we investigate and compare the statistical performance of the main modern generators we found in machine learning applications and frameworks. The use of generators, their initialization and their parallelization are a source of non-reproducibility of machine learning and artificial intelligence studies (Hill et al 2024).

For a given sequence of numbers, there are several methods for checking whether these numbers are correctly simulating randomness. Historically, Donald Knuth introduced a first series of statistical tests, published in the second volume of "The Art of Computer Programming". Despite their age, these tests remain relevant both for academic pursuit and as a reference constantly republished (Knuth, 2014). In 1995, G. Marsaglia published the first series of tests called Die Hard - the source code of which is no longer available. These tests were extended in 2005 under the name Die Harder. The NIST Statistical Test Suite (STS) is also a famous test suite particularly for assessing the ability of random numbers to be crypto-secure. A more advanced library of test suites was proposed by Pierre L'Ecuyer and Richard Simard. They released the first version of the TestU01 library in 2007 and they plan to develop a version for 64 bits generators. This library offers a wide range of tests, as well as test batteries. Among the proposed batteries, the BigCrush battery is still currently the hardest battery for generators. It is, to the best of our knowledge, the most capable test battery for evaluating the quality of PRNGs. This library also allows us to modify tests already implemented (with our own parameters) and also to add new tests. We do not use such possibilities in our study; we simply use the BigCrush battery in TestU01 version 1.2.3, which contains over 100 tests (L'Ecuyer et al., 2007).

In 1997, Makoto Matsumoto presented the Mersenne Twister (MT) generator (Matsumoto and Nishimura, 1998). This generator has shaken up the world of PRNG, and has some excellent qualities. Its outstanding features include a period of $2^{19937}$-1 and it is announced to be equidistributed in up to 623 dimensions. This generator can also be easily parallelized and delivers good quality streams (Antunes and Hill, 2023). MT is known as weak for cryptographic applications and it has some other limited flaws; his does not prevent its use in many programs or simulations. MT is still proposed as the default generator for major languages (C++, Python…) and libraries.

However, new generators came to still improve our ability to model randomness. MRG32k3a was proposed in 1999 (L'Ecuyer, 1999), Xorshift in 2003 (Marsaglia, 2003), Philox in 2011 (Salmon et al., 2011), PCG in 2014 (O'Neill, 2014) and the Xoshiro generators, which are improvements on the Xorshift generators were released recently (Blackman and Vigna, 2022) The latest generator to be tested is Xoshiro1024**, released in 2019. These last family of generators are presented as viable alternatives to MT, sometimes faster and more statistically robust (Blackman, 2021). The Philox generator is used as the default generator in TensorFlow. The aim of this study is to evaluate the statistical quality of the above-mentioned generators.

The study begins with a presentation of the methods used to test these generators. The method section will also present the resources used and the compilation methods. Next, each generator or family of generators will have a section dedicated to the presentation of its principles and the results of the batteries of tests on the various streams. The generators are then compared with each other, once again on their statistical capacities. Discussion will be provided to understand the differences between the various results obtained. The study concludes with a summary of the main results and important points. All the source code used in this study is available on a public git repository at https://gitlab.isima.fr/thwartel/testu01_variousgenerators.

## Material & Method

The evaluation of the statistical quality of the selected generators involves carrying out series of tests and then evaluating the results obtained. The generators considered are: Philox, PCG, MRG32k3a generators and some generators from the Xoshiro family (Xoshiro256++, Xoshiro256** and Xoshiro1024**). The Mersenne Twister has been intensively studied in the past, and still quite recently (Antunes et al. 2023). The tests we retain for our study are those of the TestU01 library (L'Ecuyer et al., 2007), and more specifically all the tests of the BigCrush battery. Instead of running this battery on a single initialization, we want to test and compare our generators intensively with many streams which correspond to their use in machine learning of high-performance computing. We thus use several initial states to test the generators on different streams and we try to select the different initialization states and end user would do. Each of these streams provides a result file from the BigCrush battery. The results are analyzed for potential weaknesses, and the success rates are compared with other generators and with the claims made by the PRNG creators.

In order to conduct a fair research, all generators are tested under similar conditions. For each generator, we take its official implementation, which can be found on the web page of the PRNG creator. To use the generators correctly, we follow the recommendations given by their designers. These conditions are often linked to the initialization of the generator. Thus, the approach followed for each generator initialization is as follows: if an initialization method is recommended, then we follow it. If there is no constraint, or if randomness is required to fill the generator, then we use the Mersenne Twister generator to provide this randomness. We chose to use MT because this generator is well spread and available in most modern languages. This generator is still likely to be used by scientists for many random usages. To test different initial states, we use integers as seeds in the generator or in MT. The old term seed can be misleading. It is important to realize that what is called seed often just serves an index to compute an internal initialization state of a generator. Such state is often much larger than a single integer (6 double numbers for MRG32k3a, +2 KB for MT, etc.). Like many end users accustomed to the counter generators proposed at the 2011 super computing conference, (Random numbers as easy as 1,2,3 was the best paper of this conference (Salmon et al., 2011), we use sequential indexes (seeds), the first being 0, the next seed is 1 and so on until 1000. This way, at least 1,000 streams are tested from at least 1,000 initial states. For the generators of the Xoshiro family, like Xoshiro256++, Sebastiano Vigna recommends filling in the initial state '*s'* (of 8 bytes: 256 bits) using the splitmix64 generator (Blackman and Vigna, 2022). There are no other impositions. So, we use the so-called seed, which we can vary from 0 to 1001, and we have a thousand of different initial states. This seed is placed in splitmix64, then 's' is filled. To keep track of the initializations, we write the seed (index) used and the corresponding initial state (the real seed) in the output file of each battery. These files are named BigCrush_*n*.txt where *n* represents the seed used. They contain the result of the BigCrush battery on this seed. The file names may also include High or Low to indicate the tested bits (MSB or LSB).

For each initial state, we run the BigCrush battery, necessitating between 3 to 4 hours of computing depending on the processor generation. It is important to test all the bits generated, and BigCrush is only testing 32 bits. To overcome this problem for all 64 bits generators, a common practice is to test each stream twice. The first time on the low-order 32 bits, then the second time on the high-order 32 bits. There may be several ways of using all the bits, but one method proposed by Blackman is to return the low and high parts of each number alternately. Thus, for testing a 100 of 32 bits numbers requested, we only require 50 drawings of 64 bits. We preferred to use a method that is probably closer to a real use of the generator: if the user only needs 32 bits, then he will take the low or high part and won't keep the remaining part for the next use. This is a more time-consuming method, but less cumbersome in terms of storage and implementation. The batteries are played independently, so we rely on the repeatability of the generators initialization method to guarantee that each seed gives us exactly the same initial state, and that each initial state provides exactly the same sequence of numbers. In theory, all generators should provide the same sequence of numbers from the same seed. As revealed by Antunes and Hill in (Antunes and Hill 2024), this is not always the case, and in fact depends mainly on the implementation of the initialization method and also on the platform and or framework used. Here, we're sticking to the C language, with identical implementations and we do not experienced a lack of repeatability. Some of the generators we are testing only generate 32 bits per drawing, so in order to have the same size of samples, we are testing 2002 initial states for 32 bits generators.

Each test we run produces an output file. We therefore obtain 2002 output files for 32 bits generators. The results of these tests are retrieved automatically thanks to bash scripts. Here are following information we obtain: the success rate of the streams, the number of initial states that pass or fail the battery on all bits, and the distribution of failures across the various tests. In this way, we are able to distinguish the initial states that result in streams with very good statistical qualities.

### HPC Hardware used

We have used a part of our laboratory computing cluster running Ubuntu 20.4.5 LTS and slurm-wlm 19.05.5 and gcc 9.4.0. We have been using 3 types of nodes:

1- Two AMD EPCY 7452 processors with 32 physical cores and 2 GB of RAM per core ;
2- Two Intel® Xeon® CPU E5-2670 v2 @ 2.50 GHz, 10 physical cores and 2 GB of RAM per core ;
3- Two Intel® Xeon® CPU E5-2670 0 @ 2.60 GHz, 8 physical cores and 2 GB of RAM per core.

The compilations were done with GCC on an 64-bit Intel Xeon Silver 4310 CPU @ 2.10GHz. Hyper-Threading is not activated on the cluster we used. To match real-world usage, programs are compiled using the -O2 optimization option. All the nodes are proposing a 64-bit x86 architecture.

## Results

### The Xoshiro family

In the Xoshiro generator family, we test the Xoshiro256++, Xoshiro256** and Xoshiro1024** generators. The name comes from the sequence of operations "Xor, shift, rotation". The number following the name indicates the size in bits of the generator state. This size has an impact on the generation cycle (Blackman and Vigna, 2022) but it is also suggested that it has an impact on generation quality (Vigna, 2016). These generators come with the symbol of an operator + or * that represents a "scrambler". This makes modifications to the returned value. The aim of these modifications is to improve statistical quality, + and * still show weakness in the low bits and can be considered for floating point use since the lowest bits of floating-point numbers have a smaller impact on the final double precision value. Authors advice the ** scrambler for general purpose since it allows a shuffling of all bits of the Linear Feedback Shift Register. Among other things, they prevent the systematic failure of certain BigCrush tests. Each scrambler is different and bring different features. They are fully explained in Vigna's paper on the subject, Scrambled Linear Pseudorandom Number Generator. The Xoshiro generators are in fact an improvement on the Xorshift and Xoroshiro family, which presented some statistical problems and recurring issues if the bits are processed in reverse order (Lemire et al., 2019). In this paper, we do not analyze bits in reverse order. This family of generators generates numbers with 64 bits resolution. As explained in the methods section, each stream has been tested twice (1 time for the 32 least significant bits and the other for the last most significant bits), so we've been dealing two time with the same 1001 initialization status and streams.

Figure 1 shows some statistical results obtained with these generators. It shows the number of times each BigCrush battery test failed (at least to 1 test). A distinction is made between generators, but not between high and low bits. We can see that the generators have the same weaknesses and the same strengths. They are in trouble on the latest CollisionOver tests, on ClosePairs and RandomWalks. They have relatively similar success rates, between 71.1% for Xoshiro1024** and 68.2% for Xoshiro256**. Some streams are BigCrush resistant, but not all of them so it is important to choose the right streams.  MatrixRank and LinearComp tests are rarely triggered, showing that scramblers have a significant impact on the generation quality. There's a significant difference between Xoshiro256++ and the other two: on LinearComp tests, the failure rate is almost 2 times greater for Xoshiro256++, showing that scramblers therefore have different efficiencies.

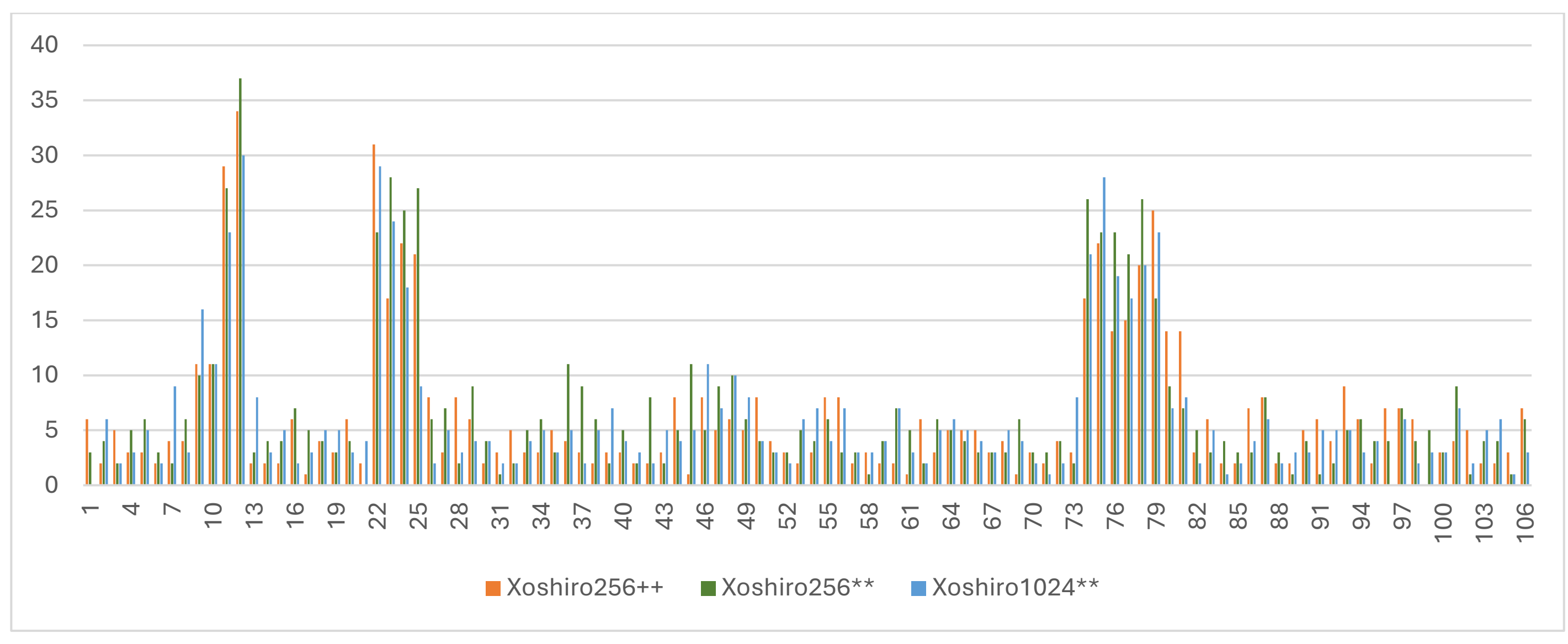

**Figure 1: Distribution of failed tests for Xoshiro256++, Xoshiro256** and Xoshiro1024** (using BigCrush test numbers)**

These results are for high and low bits combined. If we distinguish between them, high bits are more robust, with the difference between success rates being of the order of a few percent - at most 2.20% in favor of high bits for Xoshiro256**. The distribution of test failures is relatively similar for both high and low bits, across all 3 generators.

For each of the generators, we can therefore obtain high quality streams (i.e. streams that pass BigCrush on all bits). If we go back to the order of the generators on the graph, they show respectively 515, 467 and 511 high quality streams for 1001 initial states. The probabilities of obtaining such quality streams are therefore respectively 51.4%, 46.7% and 51.0%.  In the same way, there are streams that don't pass the battery on either the high or low bits. These streams are therefore completely unusable. They have 82, 101 and 88 unusable streams respectively for 1001 initial states.

**MRG32k3a**

MRG32k3a is a generator designed by Pierre L'Ecuyer, and presented in 1999. It is a multiple recursive generator (MRG). What makes its quality so good is the choice of parameters (L'Ecuyer, 1999). This generator has a 32 bits resolution, although it returns 64 bits. I couldn't find any indication of which bits to choose, so it was tested in the same way as presented by Pierre L'Ecuyer in his TestU01 user guide (L'Ecuyer et al., 2007). Given that this generator was designed by the creator of the test battery, we can suppose that the parameters were chosen with Crush resistance in mind. Even if this is the case, passing the full test battery remains a guarantee of quality. This is the oldest generator tested in our study.

The generator was tested with 2002 different initial states, due to its 32 bits resolution. For its initialization, a 384 bits initial state must be filled in (corresponding to 6 numbers in double precision with 8 bytes). This is broken down into 2 groups of 3 double-precision numbers. The combination of 6 values at 0 must be avoided. In addition, each $G_i$ group must be less than a $m_i$ value. Having no further instructions, we take the numbers and user could use (0, 1, 2, 3, ...) and place it as an index in MT, and then use 6 MT double values to fill in the states of MRG32k3a.

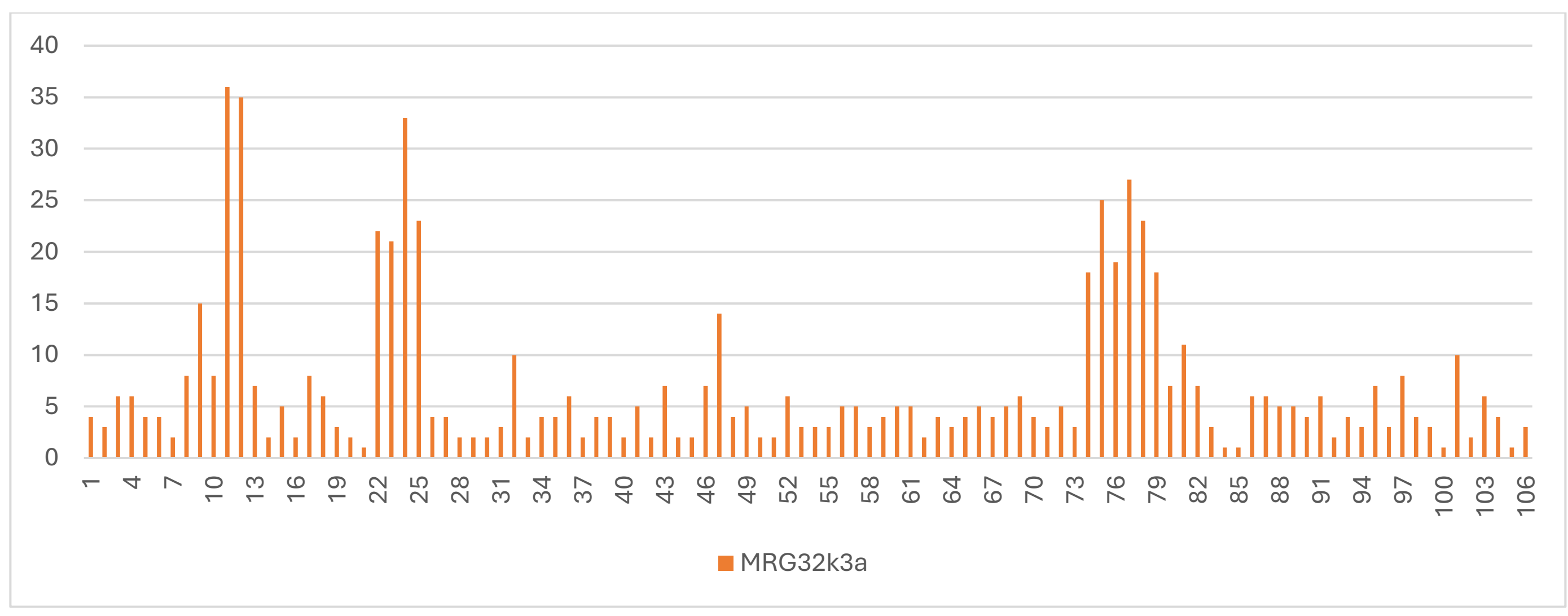

**Figure 2: Distribution of failed tests for MRG32k3a (using BigCrush test numbers)**

Figure 2 shows the distribution of MRG32k3a failures on the BigCrush battery. They were obtained using 2002 intial states. The first notable point is that there are 3 main peaks of failure. These correspond to the last CollisonOver tests, ClosePairs and RandomWalks. The maximum of test failures is less than 40 for 2002 tetst, i.e. less than 2% of occurrences. The overall success rate of the generator (success on all test in a battery) is 69.73%. Due to the 32 resolution, there is no success rate on high or low bits. Of the 2002 streams tested, 1365 are Crush resistant, and therefore are of good quality.

This generator, although old, has excellent statistical qualities. Only a small number of tests in BigCrush show some problems. Because of its rather good success rate, it can be a generator of choice if you only need 32 bits randomness and if the generation time is not a constraint since it is many times slower than generators from the Xoshiro familly. A similar version, MRG63k5a, generates numbers with a 63 bits resolution. The quality of this generator is not assessed here.

**Philox**

Philox is a generator introduced in 2011 by John Salmon. It's a counter-based generator. It has an initial state producing initial values. To obtain other values, a counter must be incremented. This generator is advertised as not being usable for cryptography (as are all the generators in this study), but it is claimed to be the fastest resistant Crush generator on GPUs (Salmon et al., 2011). It is the default generator used by the TensorFlow framework. Its counter aspect makes it very practical for parallelization: if you know the number of values required, you can easily advance through the generation cycle. This generator is available in several languages and with different resolutions. Implementations of these generators are given in the Random123 library (Salmon and Moraes, 2016).

For this study, we chose to use Philox4x32, which means 32 bits output resolution and 4 values generated in parallel by the generator. A 32 bits resolution is what TestU01 requires. Initialization of this generator consists in filling an array with 2 elements of 32 bits. We use the MT generator, in which we place the user's seed (0, 1, 2, ..., 2002) and fill the initial state with MT. Since the generator gives us a result with 32 bits of resolution, we use 2002 initial states to have as many output files as the 64 bits generators. To use the generator, you need to increment the counter for each generation. The counter is represented by four 32 bits numbers. Theoretically you can generate $4*2^{4*32}$ numbers on the same initial state. In our case, BigCrush uses a bit more than $2^{38}$ values, so we only use the first 2 numbers of the increment as they allow us to go up to $2^{66}$ values. Figure 3 summarizes the distribution of Philox failures across the 106 BigCrush tests. Once again, we can see that failures are mainly clustered on the CollisionOver, ClosePairs and RandomWalks tests. The test with the highest failure rate (test number 11) comes with 43 occurrences, representing a failure rate of 2.1%. The other tests are below 30 occurrences. The global success rate for this generator with the BigCrush battery is 69.78%. Of the 2002 initial states tested, 1397 were of high stochastic quality. Note that these states were only tested on the first $2^{38}$ values. This should be sufficient for many applications. Since Philox is sometimes available by default or implemented in a library, Philox can be considered as a very good generator for 32 bits randomness.

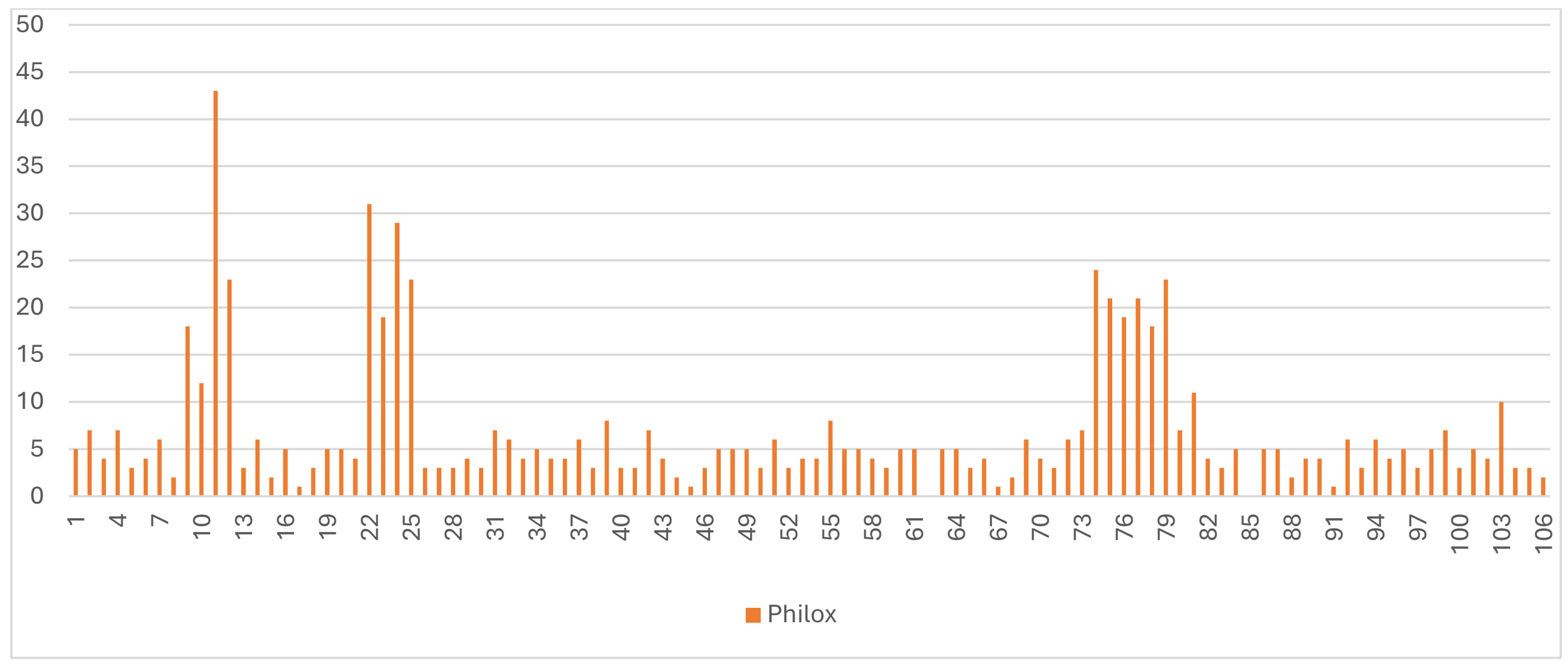

**Figure 3: Distribution of failed tests for Philox4x32 (using BigCrush test numbers)**

**PCG**

Permuted congruential generators (PCGs) were proposed in 2014 by Melissa O'Neill. This is a family claiming excellent stochastic quality as the generators are advertised as Crush resistant. This family is an evolution of the Linear Congruntial Generator (LCG) family. The quality of these generators relies on the skillful choice of coefficients, chosen by Melissa O'Neill with information from Pierre L'Ecuyer (O'Neill, 2014). As with the MRG32k3a generator, the coefficients announced as the most interesting by Pierre L'Ecuyer are perhaps chosen to pass the BigCrush tests. PCGs are easy to use, and implementations can be found on the https://pcg-random.org page with a user manual. Passing the test battery is nevertheless a guarantee of high quality. This new family claims to be better than LCG in every respect (O'Neill, 2014), knowing the structural weaknesses of LCG, this quality is highly expected. This generator was claimed to meet the needs for parallelization applications, because once the initial state has been set, an increment can be used to produce a completely random stream. In this way, from the same initial state, we can produce many streams of equal quality, and the authors expected no correlations, and this is why PCG is used in machine learning and as the default generator for the NumPy library. Unfortunately, as we mentioned in (Hill et al 2024), experience showed that PCG is weak when you use it intensively (High Performance Computing, Machine Learning, …). Thus, an extension has been proposed recently: PCG64DXSM - [https://numpy.org/doc/stable/reference/random/upgrading-pcg64.html](https://numpy.org/doc/stable/reference/random/upgrading-pcg64.html).

The bad news is that Vigna has recently shown that both are weak, not only PCG but also PCG64DXSM (https://pcg.di.unimi.it /pcg.php). In our case, we have taken the basic implementation of PCG in its minimal version. We chose a 32 bits version because this is what TestU01 requires. Higher-resolution versions are available, but they are not studied here. To use it, we need to fill in an initial state consisting of a 64 bits integer. We then select an increment and obtain a stream. To properly initialize the generator, we place our seed value (0, 1, 2, ..., 2002) in the state, set the increment to 1. Following this, due to the fact that our seed uses only a few bits, the first generated value is highly likely to be 0. To counter this, we generate one value that we discard. This generation changes the generator state and places it in a configuration that uses more bits. To obtain as many output files as for the other 32 bits generators, we use 2002 initial states on the generator.

Figure 4 shows the distribution of PCG32 failures at BigCrush battery. Like all the other generators, the failure peaks are placed on the same tests, which are: the latest CollisionOver, ClosePairs and RandomWalks tests. The tests with the highest failure rate show 33 and 32 occurrences respectively for tests 12 and 25, representing an occurrence rate of less than 1.7%. The other tests are below 30 failures. Out of the 2002 initial states tested, 1,398 streams passed BigCrush. Thanks to the 32 bits generation, each stream is tested only once. The generator has a BigCrush success rate of 69.8%. The first tested stream is not Crush-resistant. To obtain this stream, we set the generators state to 0 and then perform a warm-up to reach a state with better entropy. However, the resulting is 1, meaning it only uses 1 bit. The failure of the battery may be due to improper initialization.

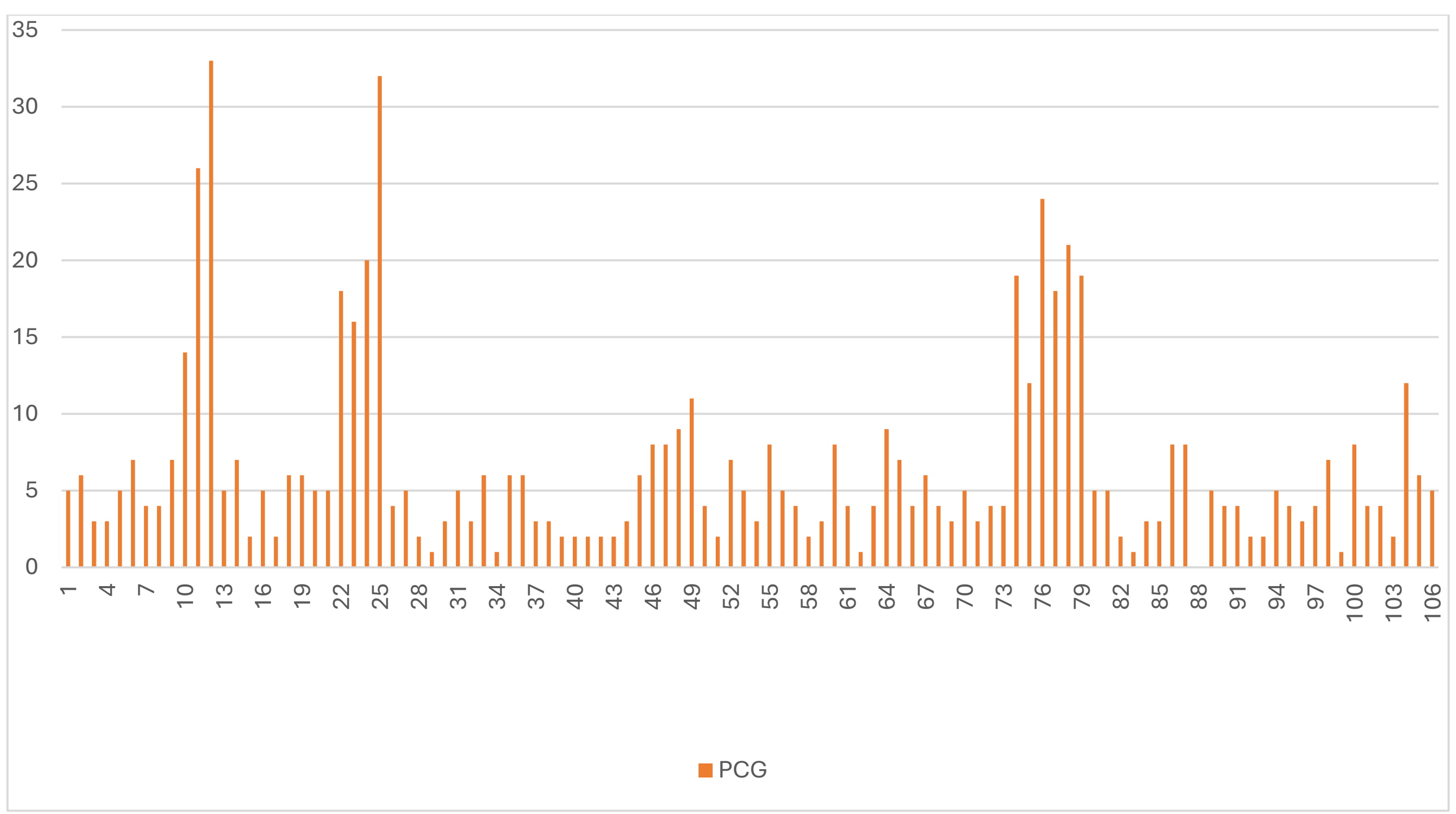

**Figure 4: Distribution of failed tests for PCG32 (using BigCrush test numbers)**

The quality of this generator in BigCrush tests makes it very interesting. However, it is important to not the issues raised by NumPy ( https://numpy.org/doc/2.0/reference/random/upgrading-pcg64.html) and Vigna (https://pcg.di.unimi.it/pcg.php) . Despite its good results, it should not be preferred. Its implementation in different languages makes it versatile. It is the default generator in NumPy, as least its version PCG64DXSM.

## Discussion

Each generator presents interesting results in the protocol we applied. However, there are a few points to note from Table 1: the 64 bits generators have a lower BigCrush battery success rate. This may be due to the fact that each stream is tested twice, once on the upper bits and then on the lower bits. This method seems more rigorous. We also note that the Xoshiro1024** generator has a larger initial state than other generators in the same family, but this doesn't give it better quality. The Xoshiro256++ generator is better than Xoshiro256**, as announced (Blackman et al., 2021).

| Generator name | Failed batteries | Overall success rate | Success rate of MSBs | Success rate of LSBs |
|---|---|---|---|---|
| Xoshiro256++ | 675 | 51.45% | 72.70% | 70.50% |
| Xoshiro256** | 733 | 46.65% | 68.80% | 67.70% |
| Xoshiro1024** | 675 | 51.05% | 71.00% | 71.20% |
| Philox | 605 | 69.78% | / | / |
| MRG32k3a | 606 | 69.73% | / | / |
| PCG32 | 603 | 69.83% | / | / |

**Table 1: Some generator statistics**

If the user needs a 64 bits generator, Xoshiro256++ could be the best choice. It is the one with the best statistics. In the case of a 32 bits generator, this is the one with the best results. High bits are the most robust. However, this generator has a higher cost, uses a larger initial state than Philox and PCG, and its use will produce 64 bits where only 32 are needed, so there may be some loss. This is why the Philox generators is a perfectly coherent alternative – PCG could also be considered purely based on our results. So is MRG32k3a, but the initial state is very large (384 bits), compared with 64 bits for Philox4x32 and PCG32. In terms of ease of use, Philox should be prioritized, as implementations is available in default libraries and user functions are provided with them.

**Comparison between Philox and PCG**

Philox and PCG are both 32 bits generators, implemented in libraries and used for parallelization. It therefore makes sense to compare these generators directly. Table 1 already shows battery success rates in favor of PCG (difference between the 2 of 0.005% over 2002 tests). PCG has passed only 2 more tests.

Figure 5 shows the distribution of Philox4x32 and PCG32 generator failures on the BigCrush test battery. The same 3 failures peaks can be seen, as well as a striking similarity between the 2 generators. Each test has a similar number of failures.

A difference can be made on the test numbers 11 and 12. Both are CollisionOver tests, but with different parameters (L'Ecuyer and Simard, 2007). Test number 11 was more successful for PCG, with 26 failures versus 43 for Philox. On the test number 12, the opposite is true, with Philox doing better with 23 failures versus 33 for PCG. So, the same test but with different parameters is better passed by one or other of the generators.

These 2 generators produce extremely similar results. They should be tested on more initial states, so as to have more samples to analyze. It would also be necessary to test other characteristics expected on good PRNG to differentiate them. We can mention the generation speed, the cycle length, the ability to parallelize properly and the memory footprint.

In her paper, Melissa O'Neill emphasizes that passing a statistical test suite like TestU01's BigCrush does not guarantee a high-quality random number generator (RNG). Two RNGs might both pass BigCrush, yet one could be significantly weaker—either because the test isn't sensitive to its flaws or because it relies on an unusually large internal state. In fact, all finite-state RNGs, even ideal ones, will eventually fail if tested exhaustively, simply because they have a limited amount of randomness.

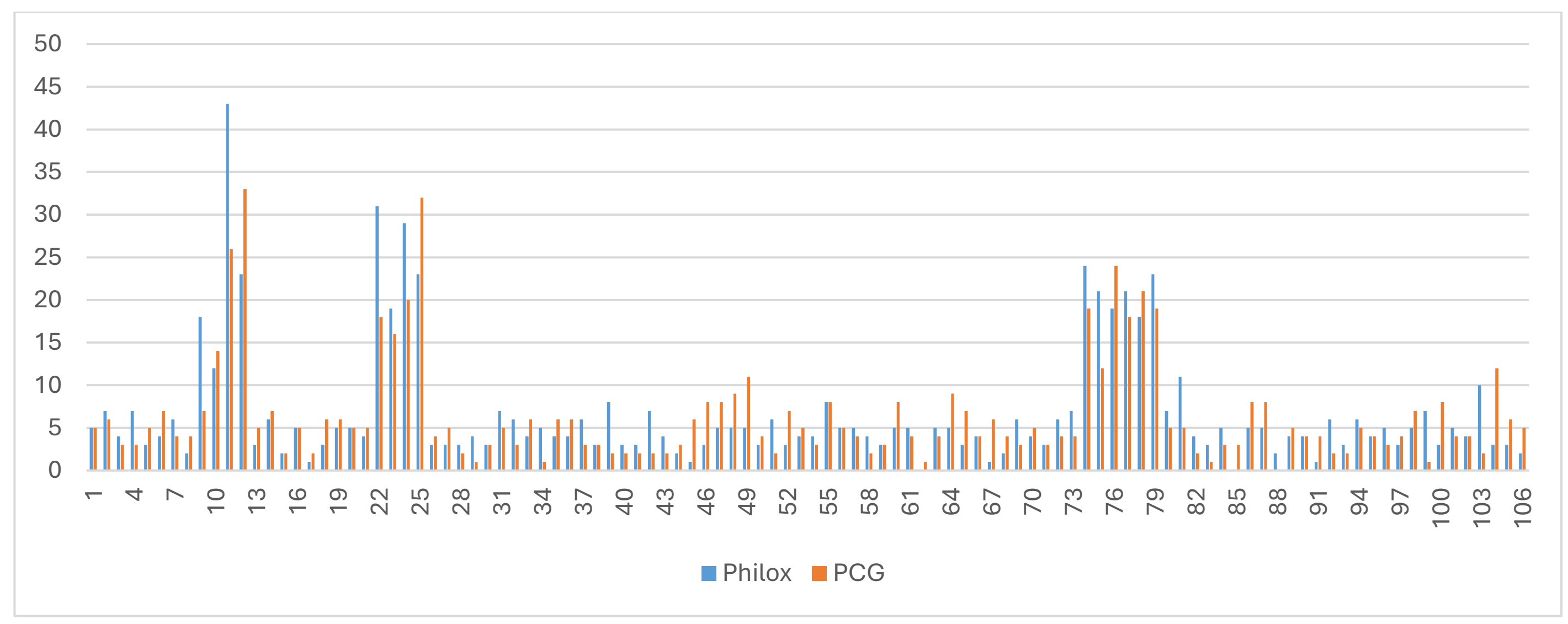

**Figure 5: Distribution of failed test for Philox4x32 and PCG32 (using BigCrush test numbers)**

For example, an RNG with 32 bits of internal state producing 32-bit outputs can only emit each possible value once per cycle, eliminating repeated values—a natural aspect of randomness. This behavior contradicts expectations, as illustrated by the birthday paradox: in a group of just 23 people, there's a 50% chance two share a birthday. True randomness allows for repetition. When an RNG's state is too small to permit such repeats, its output appears artificially uniform. To address this, O'Neil introduces the concept of headroom, which measures how much larger a generator's state is than the minimum required to pass a given test. For example, BigCrush is estimated to need at least 36 bits of state: a generator with 128 bits thus has 92 bits of headroom, while one with just 40 bits is exceptionally efficient. Estimates suggest SmallCrush requires ~32 bits, Crush ~35 bits, and BigCrush ~36 bits. Generators passing with minimal state are noteworthy, while those needing far more may be statistically inefficient despite passing.

From this perspective, BigCrush failures in generators with fewer than 36 bits of state likely reflect insufficient capacity rather than poor quality. However, all generators examined in our study have at least 64 bits of state, implying they should have adequate headroom to pass BigCrush. Failures observed in these generators, therefore, likely stem from deeper statistical flaws or issues in implementation that warrant closer analysis.

It's also critical to interpret BigCrush results in light of multiple testing effects. BigCrush includes around 160 individual tests. Under the null hypothesis that an RNG is statistically sound, p-values should be uniformly distributed between 0 and 1. The probability of any single p-value falling below 0.001 or above 0.999 is 0.002. But with 160 tests, the chance of seeing at least one such extreme value rises to approximately 27.4%. This means that even high-quality RNGs may fail one or more tests by chance alone. Thus, isolated test failures shouldn't be overemphasized; it's the pattern and consistency of failures that indicate true issues.

In our analysis, if we apply a conservative filter—excluding only the most extreme p-values (less than $10^{-15}$, reported as 1-eps by BigCrush)—to avoid overreacting to anomalies, we obtain that all the PRNGs in this study (expect MT), are not failing any BigCrush tests, for all the 2002 random streams. Considering the Crush-Resistance only with systematic failure, or with failure having a p-values as extreme as $10^{-15}$, is subject to discussion.

It might be necessary to interpret p-values in the context of multiple testing. Even an ideal RNG is expected to produce extreme p-values (outside the 0.001–0.999 range) roughly 0.2% of the time. Over 160 tests, this results in a 27.4% chance of at least one such occurrence. To avoid misinterpretation, when excluding only the most extreme cases ($p < 10^{-15}$), all the PRNGs provided by authors seems to be passing all BigCrush tests over 2002 random streams.

## Conclusion

Good quality randomness is a prerequisite for many research applications. With this in mind, we have analyzed 6 generators that are widely used and that can reach excellent statistical quality. All the generators are advertised as Crush-resistant, and they can all pass at least one BigCrush battery. However, this claim is not true, and this paper shows that it is easy to find initial states where all generators fail to pass all the tests of all BigCrush. We were also interested in the distribution of the failed tests. The PRNG authors rarely document the the initial state they used to pass all the tests, thus limiting reproducibility. This other input of this paper is that we give a BigCrush success rate. We determined these success rates by running each generator through more than 1000 BigCrush tests with different streams. Interestingly, this approach shows that the different generation methods produce similar results. Between the strongest and the weakest generator of this study – the difference is very small: 2.16% of difference in success rates (if only 32 bits are used). The generation methods used in this test are distributed over 20 years, we could have expected a quality increase, and this was not the case.

All tested generators were far from 100% successful, and top generators were around 70% of success. We would like to be able to use a source of pure randomness (storing a 1000 times $2^{39}$ values and see how it performs against the BigCrish battery. We consider these 6 generators are of sufficient quality to be used in sequential many programs. For parallel applications, each use should make a particular care to initializing the generator with streams that were completely successful to the BigCrush battery. A side remark that we must recall is that a seed is not more equivalent to an initial state for modern generators. In highly sensitive simulations, it will be necessary to select only the streams that appear to be of the best quality. The PCG generator has shown good results, but we do not recommend its use.

To complete this study a few points can be umproved. As a first step, it would be interesting to test other 64 bits generators and compare them with the Xoshiro family, to see whether their success rate is naturally below that of 32 bits generators, or whether they are worse.We could also test the MT generator to use it as a reference value for the others. In addition, in order to compare the Philox and PCG generators, which are intended for identical uses, we could need to run more batteries on each of them and compare Philox with the PCG64DXSM version. For all generators, we would also need to evaluate their generation speed ans the memory space they occupy. It would also be interesting to test -O3 optimization on these generators ans study variations in their behavior.

## Web Reference

## Biographies

**Theau Wartel** is a student at the ISIMA - Clermont Auvergne INP engineering school. He specializes in embedded and virtual interactive systems. He is part of the “research course”, which involves students and researchers working together. He has carried this research in the LIMOS laboratory (UMR CNRS 6158).

**David R. C. Hill** is a full professor of Computer Science at University Clermont Auvergne (UCA) doing his research at the French Centre for National Research (CNRS) in the LIMOS laboratory (UMR 6158). He earned his Ph.D. in 1993 and Research Director Habilitation in 2000 both from Blaise Pascal University and later became Vice President of this University (2008-2012). He is also past director of a French Regional Computing Center (CRRI) (2008-2010) and was appointed two times deputy director of the ISIMA Engineering Institute of Computer Science – part of Clermont Auvergne INP, #1 Technology Hub in Central France (2005-2007 ; 2018-2021). He is now Director of an international graduate track at Clermont Auvergne INP. Prof Hill has authored or co-authored more than 300 papers and has also published several scientific books. He recently supervised research at CERN in High Performance Computing (https://isima.fr/~hill/).

## Roles